\documentclass[
aps,%
12pt,%
final,%
notitlepage,%
oneside,%
onecolumn,%
nobibnotes,%
nofootinbib,%
superscriptaddress,%
noshowpacs,%
centertags]%
{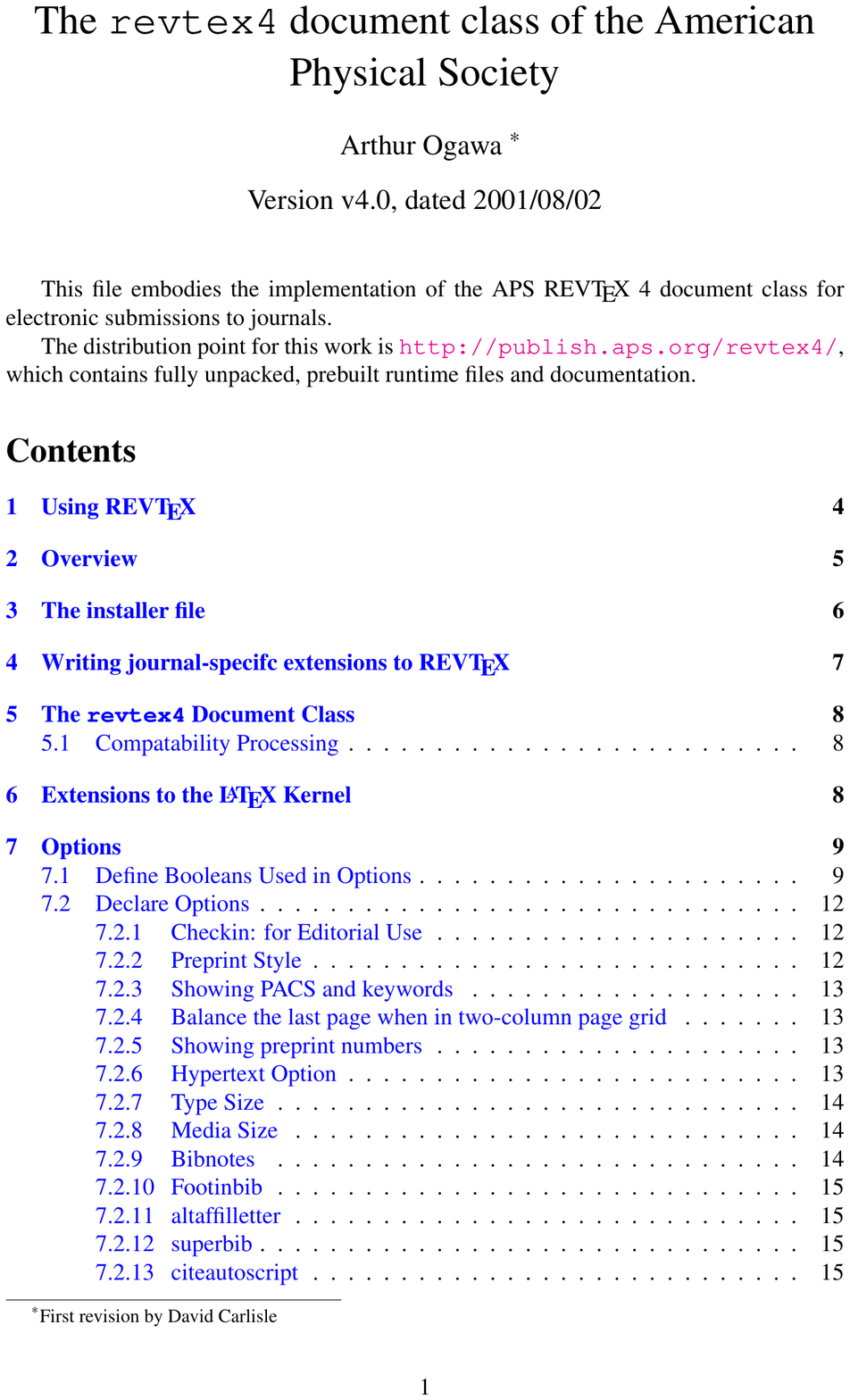}

\usepackage[german,british]{babel}
\usepackage{amssymb,amsmath}
\usepackage{graphicx}

\begin{document}

%
\title{Search for Gluinos with the CMS Detector at the LHC}

\author{\firstname{\copyright{}E.} \surname{Ziebarth}}

\email[]{ziebarth@cern.ch}
\affiliation{ Institut fuer Experimentelle Kernphysik, Universitaet Karlsruhe, Karlsruhe, D-76136}

\begin{abstract}
Gluino pair production in an EGRET motivated supersymmetry scenario has been studied. The analysis shows, that this special scenario could be proven with the first LHC data, and assuming a higher integrated luminosity the analysis would be able to cover a large region of the minimal supergravity (mSUGRA) parameter space.
\end{abstract}

\maketitle


\section{Introduction}
\begin{figure}[ht]
\caption{Constraints on the mSUGRA parameter space for $\tan \beta=50$.  The region, which is allowed by the results of WMAP are assigned in blue. The EGRET data require low $m_{1/2}$ values, which implies values of $m_0$ in the tera-electron-volt range, if all constraints from the Higgs limit ($m_h$), $g-2$ and the b $\rightarrow$ s$\gamma$ branching ratio are considered (limits exclude regions to the left). The CMS point LM9 ($m_0=1450$, $m_{1/2}=175$) is located in the region which is preferred by theory and experiments\label{FigParamSpace} \cite{CERNCour}.}
\bigskip
\includegraphics[width=11cm]{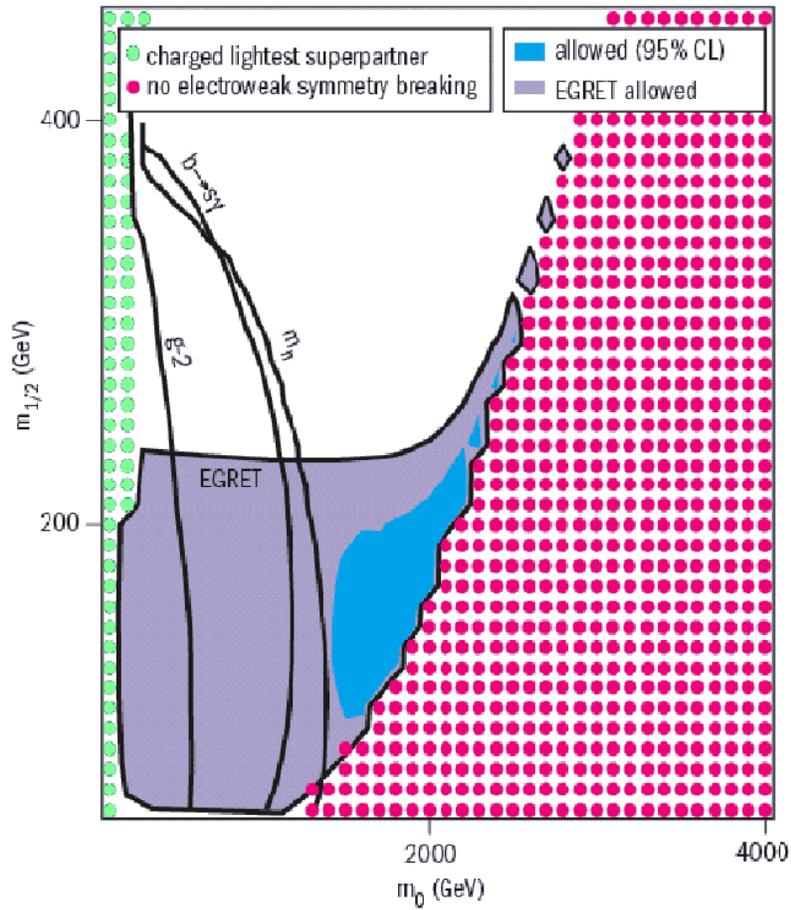}
\end{figure}

\begin{figure}[ht]
\caption{Cross sections of SUSY processes at the CMS parameter point LM9 normalised to the sum of all SUSY cross sections versus the centre of mass energy of the collider. At the Tevatron (about 1000 GeV) neutralino-chargino pair production $\tilde{\chi}^{\pm}\tilde{\chi}^0$ would be the dominant process, while the dominant process at the LHC (14000 GeV) will be gluino pair production ($\tilde{g}\tilde{g}$). All processes, which are not plotted (like squark pair production $\tilde{q}\tilde{q}$), contribute less than 10 to the SUSY cross section.\label{FigCSLM9}}

\bigskip
\includegraphics[width=16cm]{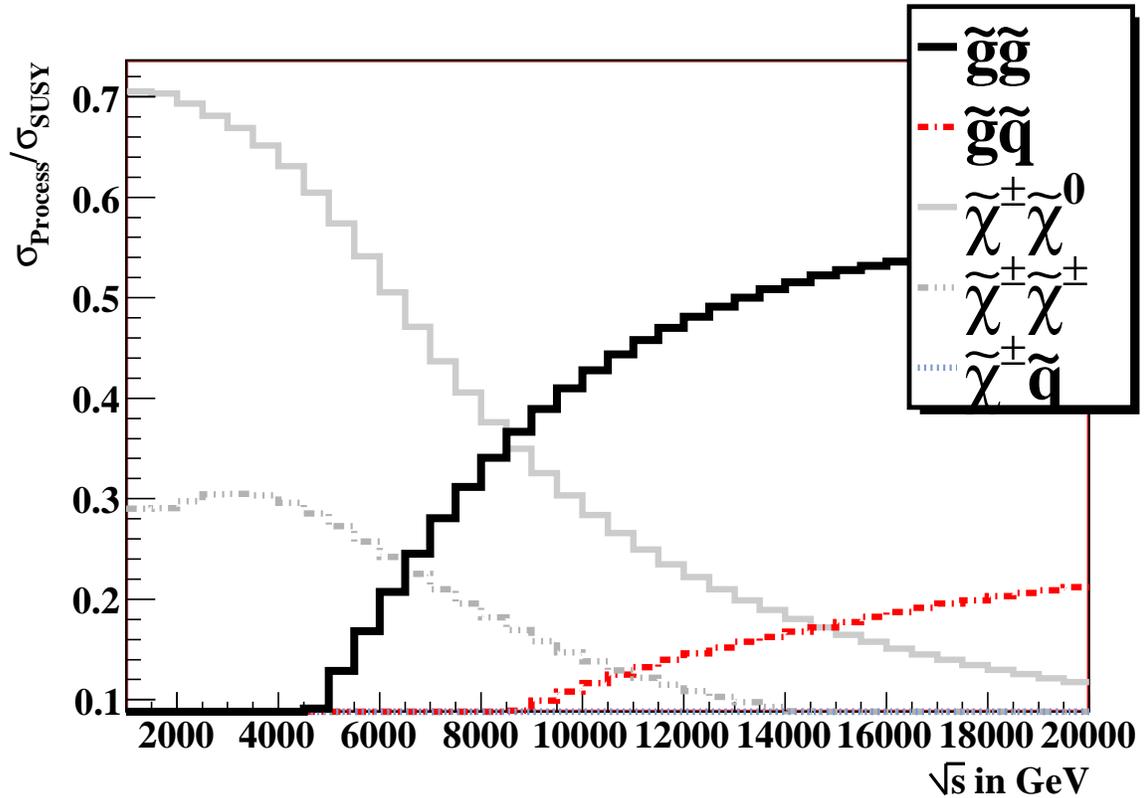}
\end{figure}

\begin{figure}[ht]
\caption{Feynman diagrams for gluino pair production. Gluino pairs can be produced from a gluon directly (like plotted in the first line) or via exchange of a gluino or a squark (like plotted in the second and third line). In the LHC production from quark pairs (right row) as well as production from gluon pairs (left row) contribute to the production cross section.\label{FigGluinoProd}}
\bigskip
\includegraphics[width=16cm]{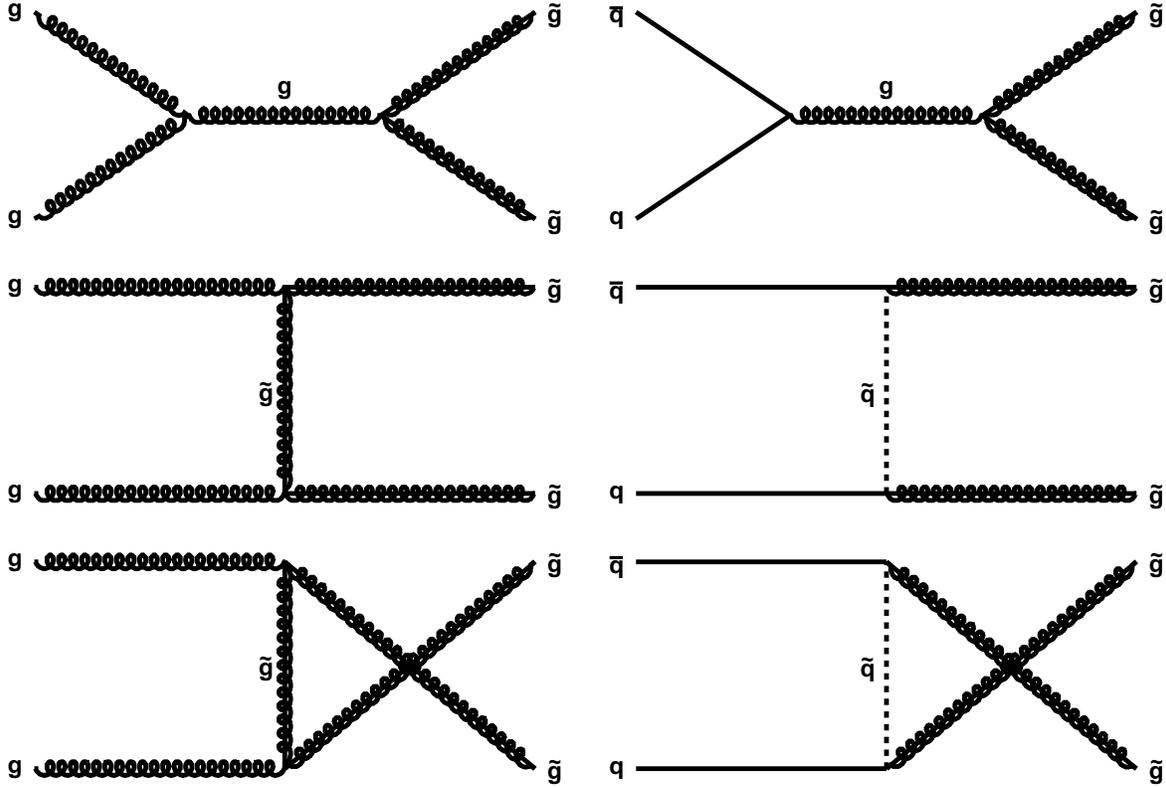}
\end{figure}

\begin{figure}[ht]
\caption{Feynman diagrams for gluino decays at LM9 and in the focus point region. For LM9 the first diagram is strongly preferred compared to the other ones. In the focus point region, at low scalar masses the branching ratio of the second digram gets two times larger than the one of the first diagram. At higher fermion masses (thus also higher scalar masses) the channels, which are plotted in the lower line, are open.\label{FigGluinoDec}}
\bigskip
\includegraphics[width=8cm]{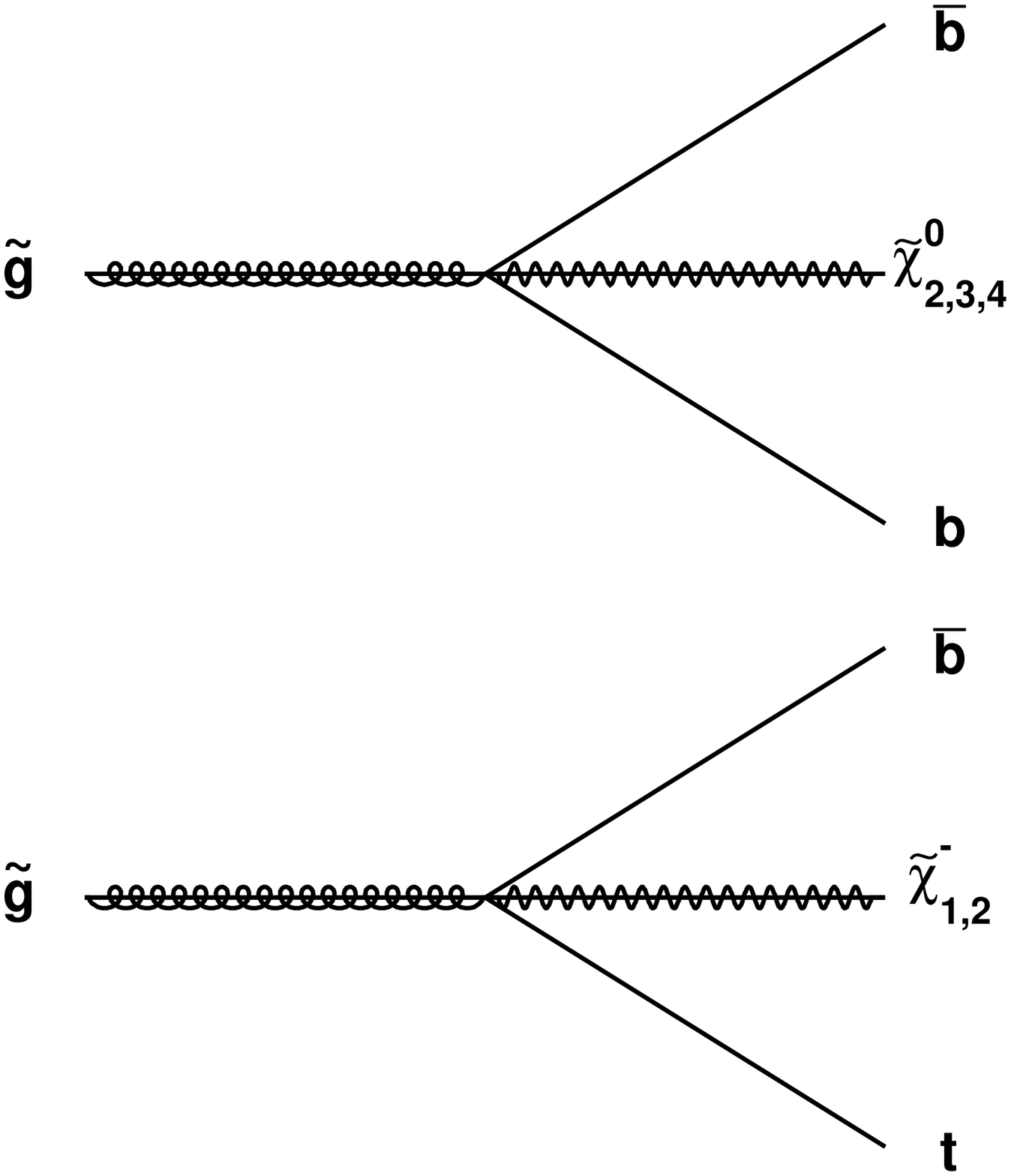}
\includegraphics[width=8cm]{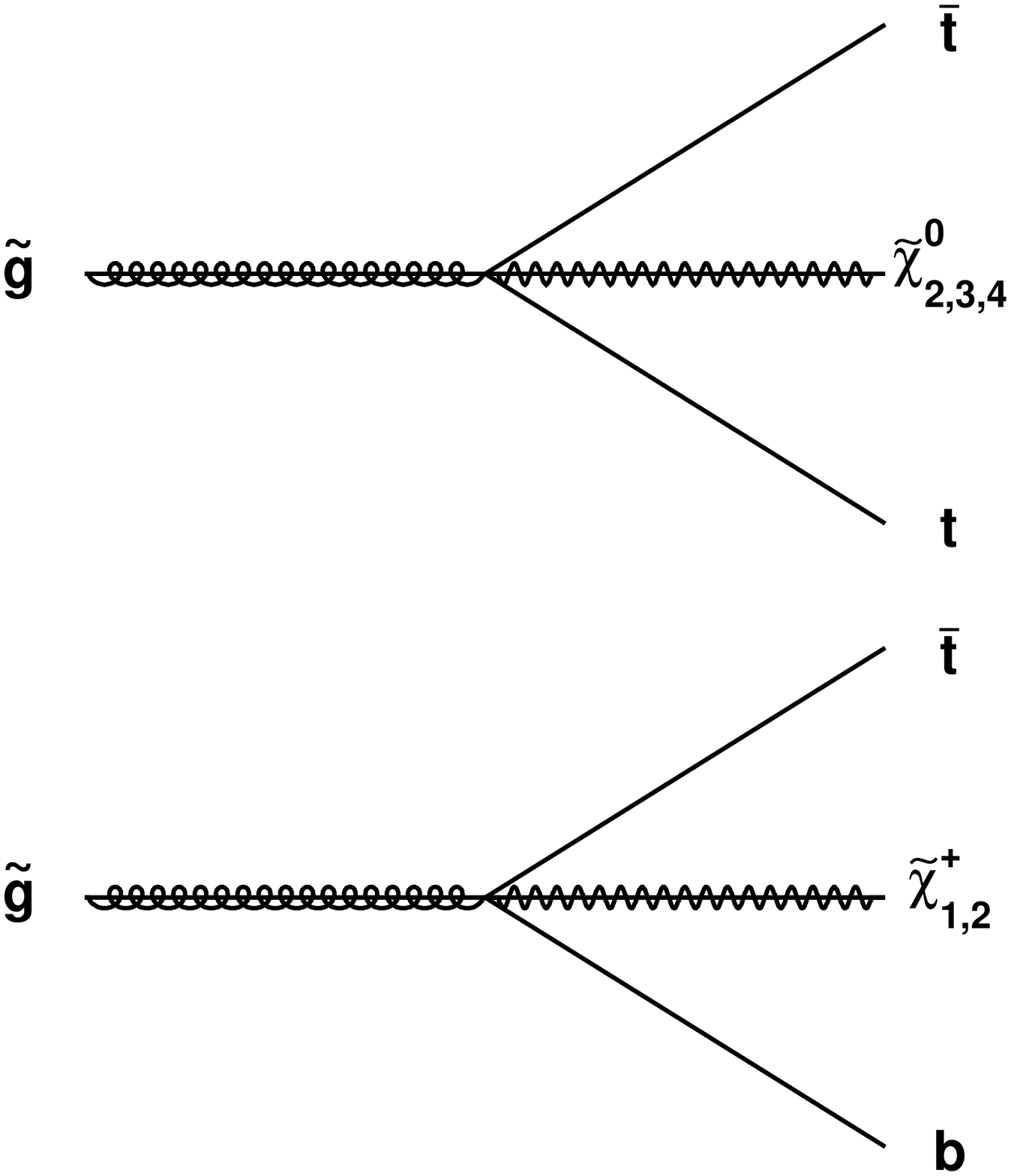}
\end{figure}

\begin{figure}[ht]
\caption{Two examples for decay cascades at LM9 and in the FPR. At LM9 the decays to top quarks are strongly suppressed, because the scalar mass parameter is very low. The neutralino results in missing transverse energy, because it is electrically neutral and weakly interacting. Leptons contribute to the lepton number and jets to the jetmultiplicity as well as during their transverse momenta to the effective mass. \label{FigCascades}}
\bigskip
\includegraphics[width=16cm]{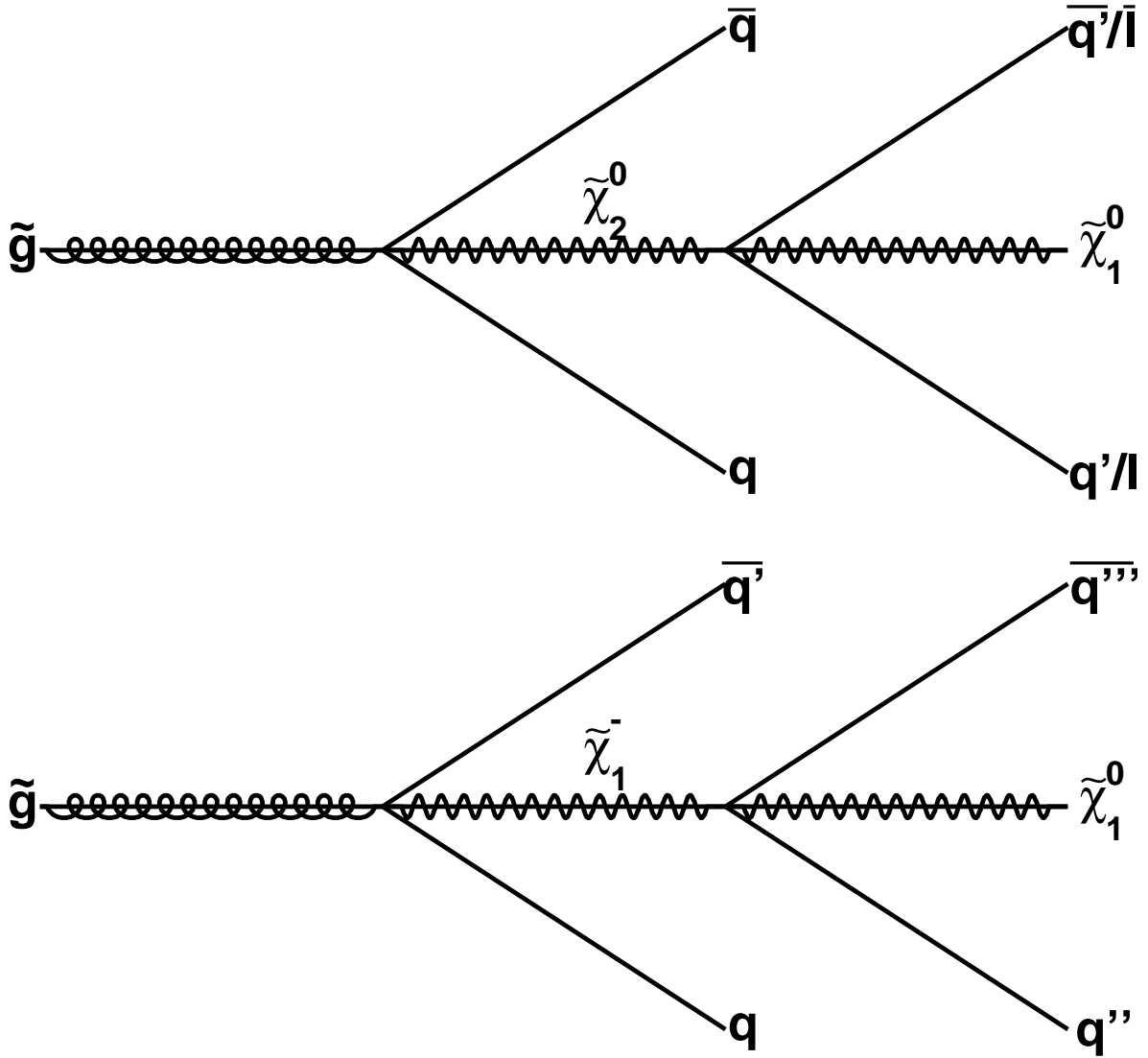}
\end{figure}

The Minimal Supersymmetric Standard Model (MSSM) is an extension of the standard model including a symmetry between fermions and bosons. Such models introduce fermion partners to all standard model particles and bosonic ones to all fermions (SUSY particles). Assuming, that typical SUSY masses are located  at the TeV scale would make the unification of the coupling constants easily accessible \cite{Unification}. Additionally, in R-parity conserving models the neutralino would provide a candidate for dark matter, a so-called WIMP (weakly interacting massive particle) \cite{EGRET}.\\
Due to the fact, that supersymmetry (SUSY) has not been discovered yet, high masses for SUSY particles can be achieved by applying Soft Symmetry breaking mechanisms \cite{Primer}. These symmetry breaking mechanisms results in general in more than 100 free parameters, which are reduced to five in minimal supergravity (mSUGRA) models: the common scalar (fermion) mass parameters $m_0$ ($m_{1/2}$) at the GUT scale, the sign of the Higgs mass parameter $\mathrm{sign}(\mu)$, the trilinear coupling $A_0$, and the ratio between the two neutral Higgs vacuum expectation values of the two Higgs doublets$\tan \beta$ \cite{Primer}. Thus the SUSY parameter set is given by
\begin{equation}
 	\lbrace p \rbrace =  \lbrace \mathrm{sign}(\mu), m_0, m_{1/2},  A_0,  \tan \beta \rbrace. 
\end{equation}
Due to the fact, that the masses of fermions, hence their cross sections and decay channels, strongly depend on the chosen mSUGRA parameters an analysis cannot cover the whole mSUGRA parameter space. Thus a reasonable region for the analysis should be chosen. This choice can be motivated by the EGRET experiment, with an excess of high energetic galactic gamma radiation has been observed. This excess can be explained by dark matter annihilation with neutralino mass in the range 50 to 70 GeV \cite{EGRET}. This condition corresponds to a special regions in the mSUGRA parameter space:
The region at low fermion masses, which corresponds to low $m_{1/2}$. Here the mass of the neutralino is given by $m_{\tilde{\chi}}\approx0.4 m_{1/2}$, while the mass of gluinos is $m_{\tilde{g}}\approx2.7 m_{1/2}$. In this region the masses of $\tilde{\chi}_2^0$ and $\tilde{\chi}_1^{\pm}$ are about one fourth of the gluino mass. The value for the dark matter density given by these models would be consistent with the results of WMAP \cite{EGRET}. The regions, which are preferred by experimental and theoretical results are shown in Fig. \ref{FigParamSpace} \cite{CERNCour}.\\
Another region, which is consistent with the results of WMAP is the focus point region (FPR) \cite{FPR}. This region is a narrow band along the region, which is forbidden by electroweak symmetry breaking. In this region the masses of the two lightest neutralinos get very close to each other, which makes them independent of $m_{1/2}$. The mass of the gluino is expected to be about one order of magnitude larger than the ones of$\tilde{\chi}_2^0$ and $\tilde{\chi}_1^{\pm}$.\\
Thus the cross sections for gaugino production are expected to be higher compared with the cross sections of sfermion production. The production channels with the largest cross sections in this scenario are expected to be gluino pair $\tilde{g}\tilde{g}$ production and direct production of neutralino chargino $\tilde{\chi}_2^0\tilde{\chi}_1^{\pm}$ pairs, which can be detected as a trilepton signature \cite{Trilepton}. The relation between the two production channels can be determined from the mass hierarchies, as they have been discussed before. In the low $m_{1/2}$ region the cross section for gluino pair production is expected to be one order of magnitude higher than the one for neutralino chargino pairs, which is shown in Fig. \ref{FigCSLM9}, while in the FPR neutralio-chargino pair production is preferred.\\ 
The trilepton signature is hard to separate from the background. Thus it will be harder to detect than gluino pair production. Additional, the decay cascades of gluino production are expected to be longer, thus more parameters can be determined from the same data.\\
Gluino pair production is in both regions the gluino production channel with the highest cross section. The Feynman diagrams, which contribute to gluino pair production are given in Fig. \ref{FigGluinoProd} \cite{Primer}. In the low $m_{1/2}$ region the production cross sections are large compared to other regions (in the 10 pb range) caused by the low mass of the gluino. In the focus point region the gluinos are much heavier, thus the production cross sections vary between pb and fb.\\
In both regions the gluino is lighter than any squark. This results in three-body decays via virtual squarks. The products continue decaying into lighter SUSY particles until the LSP is reached. Thus the length of the decay chain depends on the mass of the gluinos and the gauginos. 
The branching ratios of the decays to bottom and top quarks are strongly related to the initial values of the SUSY breaking. For the low fermion mass region the decay to bottom quarks is strongly preferred. Thus the main decay channel is the decay into a neutralino and a bottom pair and the decay into a bottom quark. Caused by the low mass of the gluino the decay cascades appear not to be very long. In the lower part of the focus point region also the decay to top quarks contributes and in the higher part also to channel including a bottom, a top and a chargino is open. The main contributing Feynman diagrams for gluino decays are plotted in Fig.~\ref{FigGluinoDec}. Two examples for cascades at LM9 and in the FPR are given in Fig.~\ref{FigCascades}.\\
\section{Analysis}
\begin{figure}[ht]
\caption[Number of jets versus number of leptons]{Number of jets versus number of leptons. The event numbers have been normalised to one. Due to jet missreconstruction the number of jets gets similar to the one of gluino pair production. As result the topology of $t\bar{t}$ production is very similar to gluino pair production. \label{nJetsnLep}}
\bigskip

\includegraphics[width=16cm]{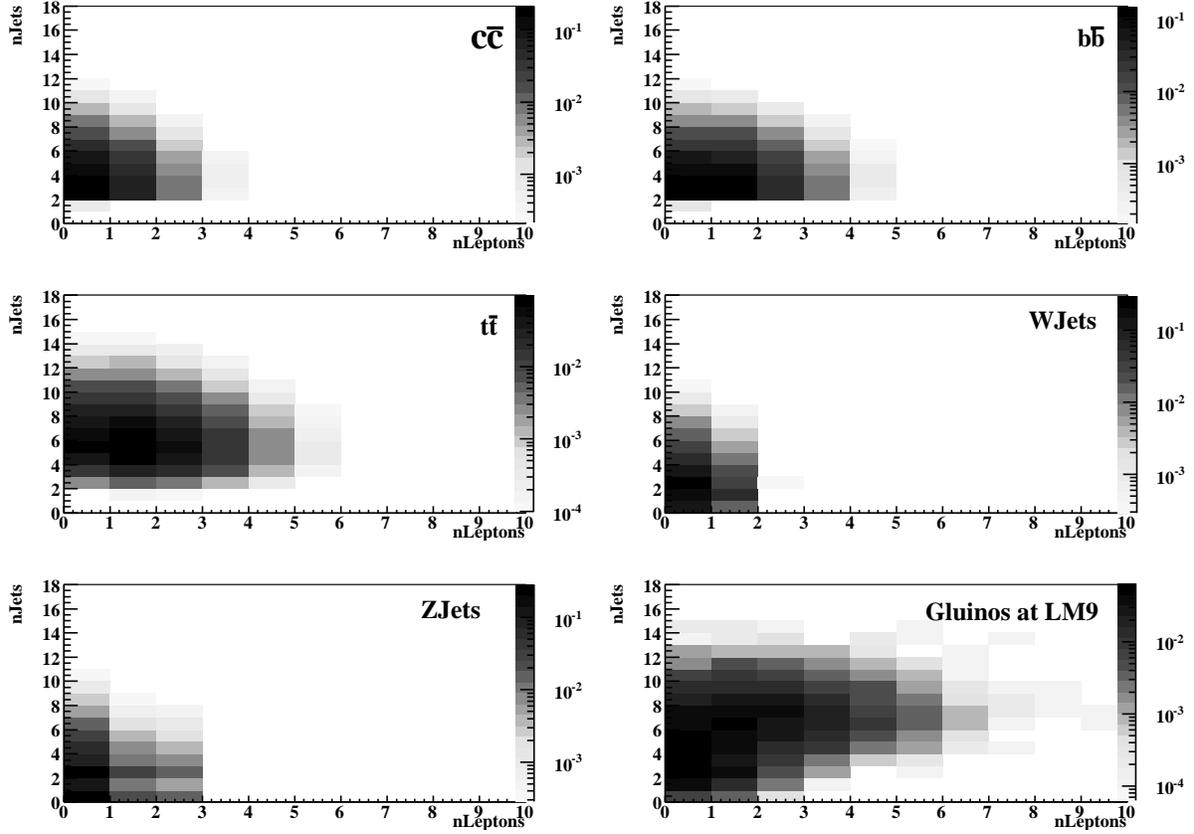}
\end{figure}

\begin{figure}[ht]

\caption{MET versus number of jets. The event numbers have been normalised to one. The MET is very different for $t\bar{t}$ events and gluino pair production. That is because the lightest supersymmetric particle, which is very heavy, is escaping from the detector. The MET can be used very well to distinguish between signal and background. \label{nJetsMET}}
\bigskip
\includegraphics[width=16cm]{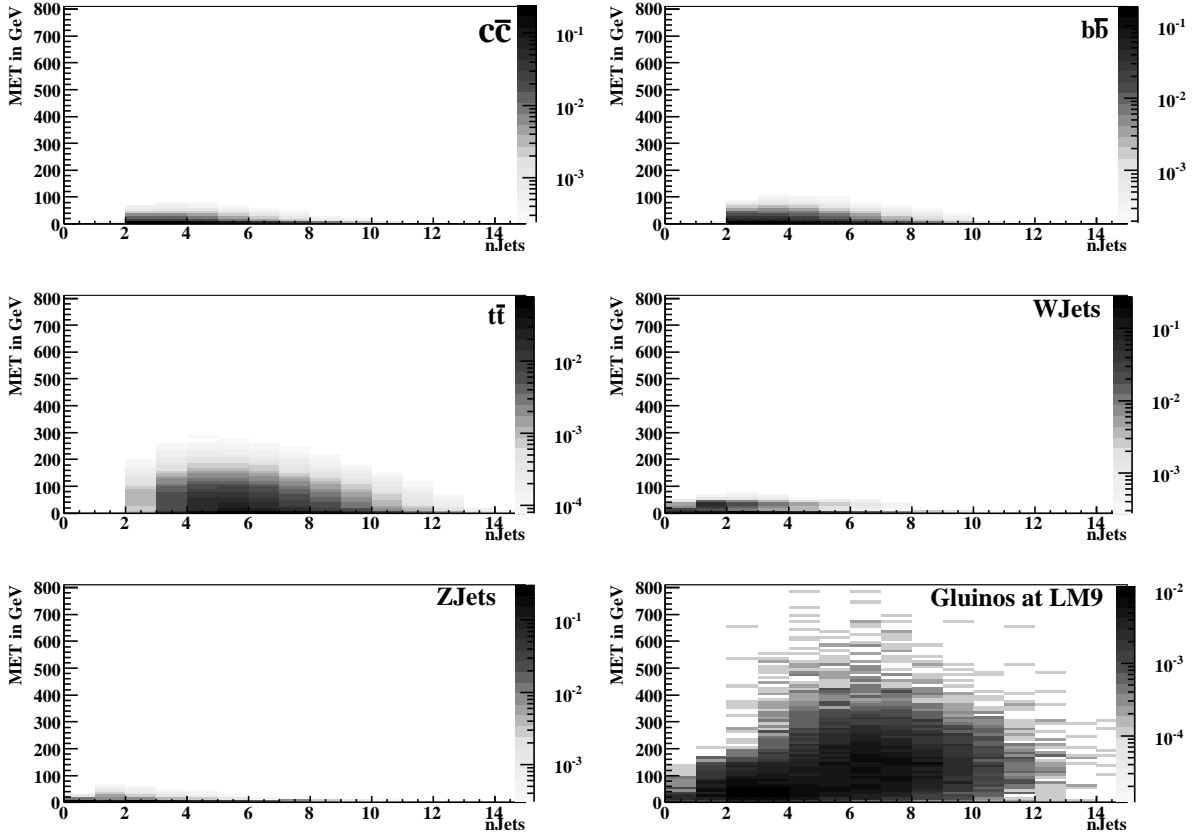}
\end{figure}

\begin{figure}[ht]
\caption{Scattering plots including two dimensional cut function. The scattering plots are normalised to one, thus the ones of the backgrounds include a lot more events. The black lines assign the two dimensional cuts, which have been applied to check the consistency of the neural network analysis. The grey lines show the precuts, which have been done to prepare the data samples for the neural network training.\label{Fig2DCuts}}
\bigskip
\includegraphics[width=7.9cm]{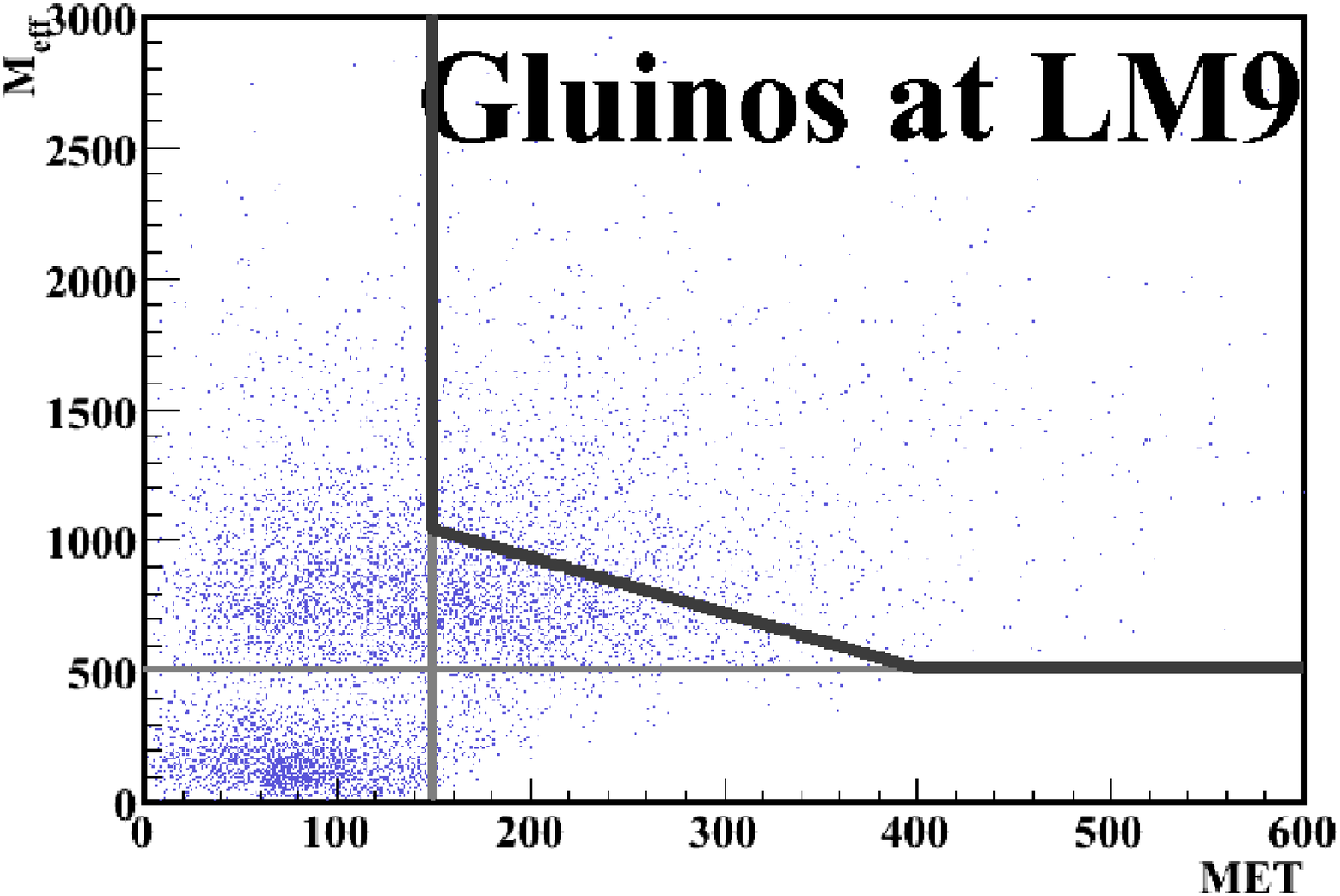}
\includegraphics[width=7.9cm]{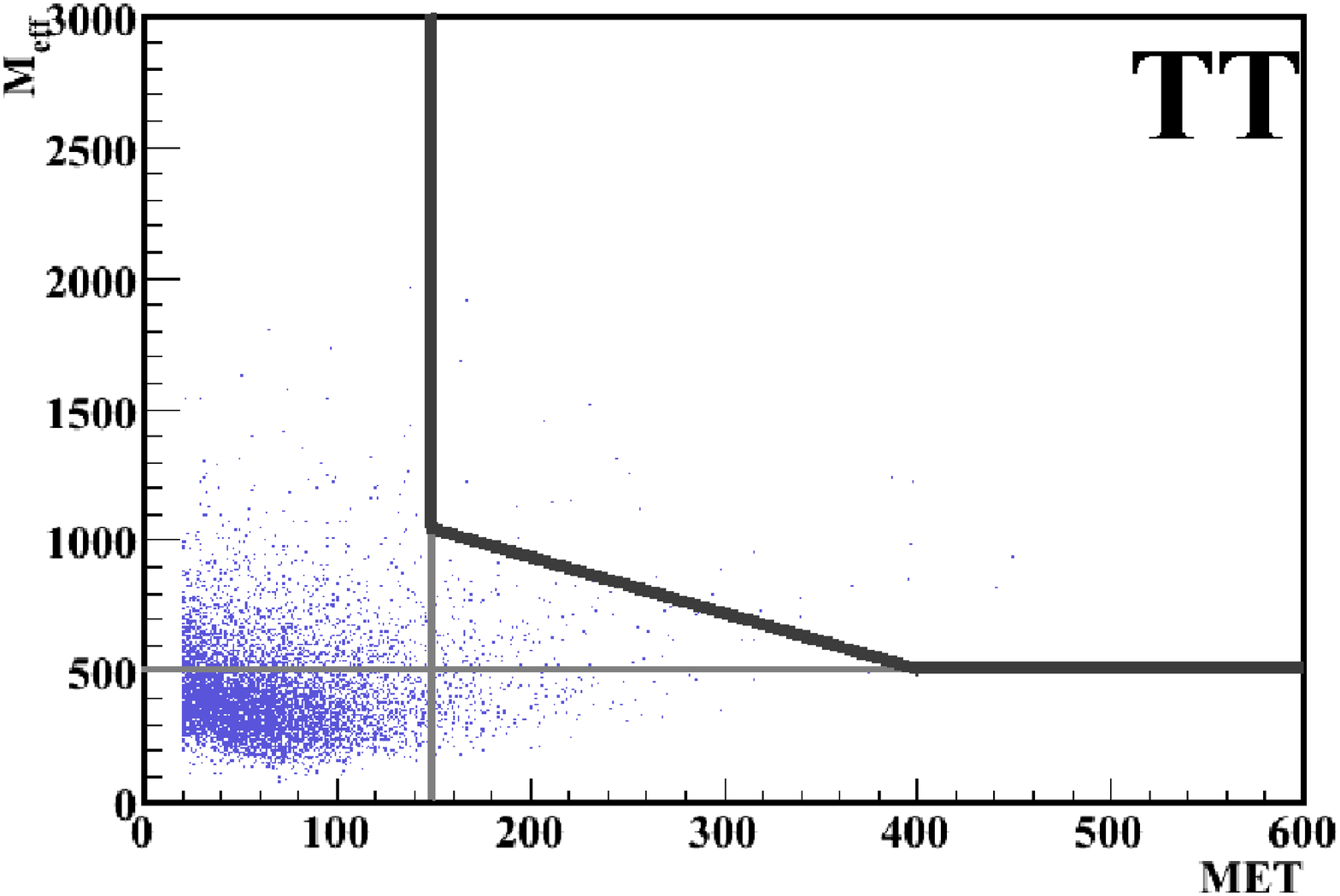}\\
\includegraphics[width=7.9cm]{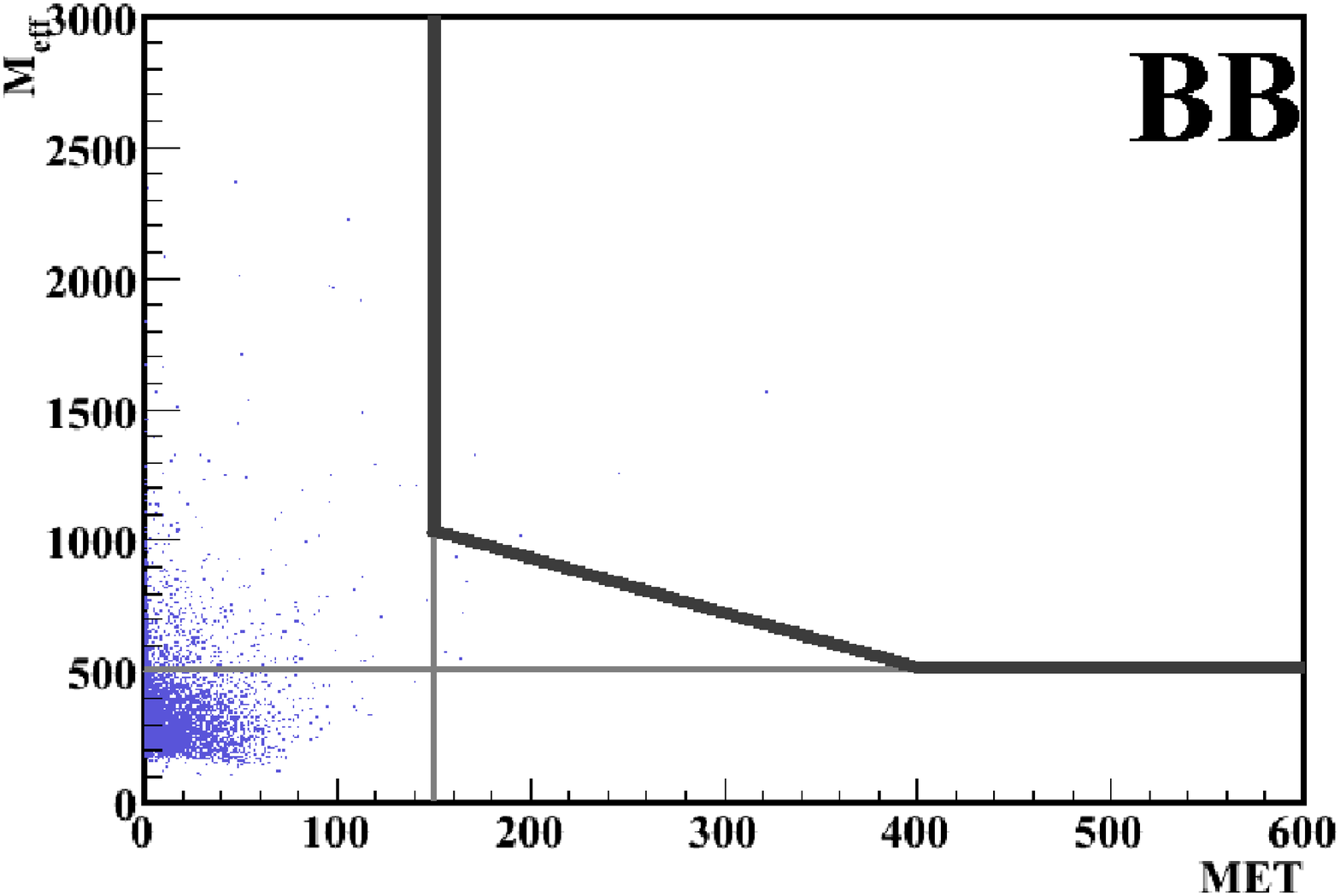}
\includegraphics[width=7.9cm]{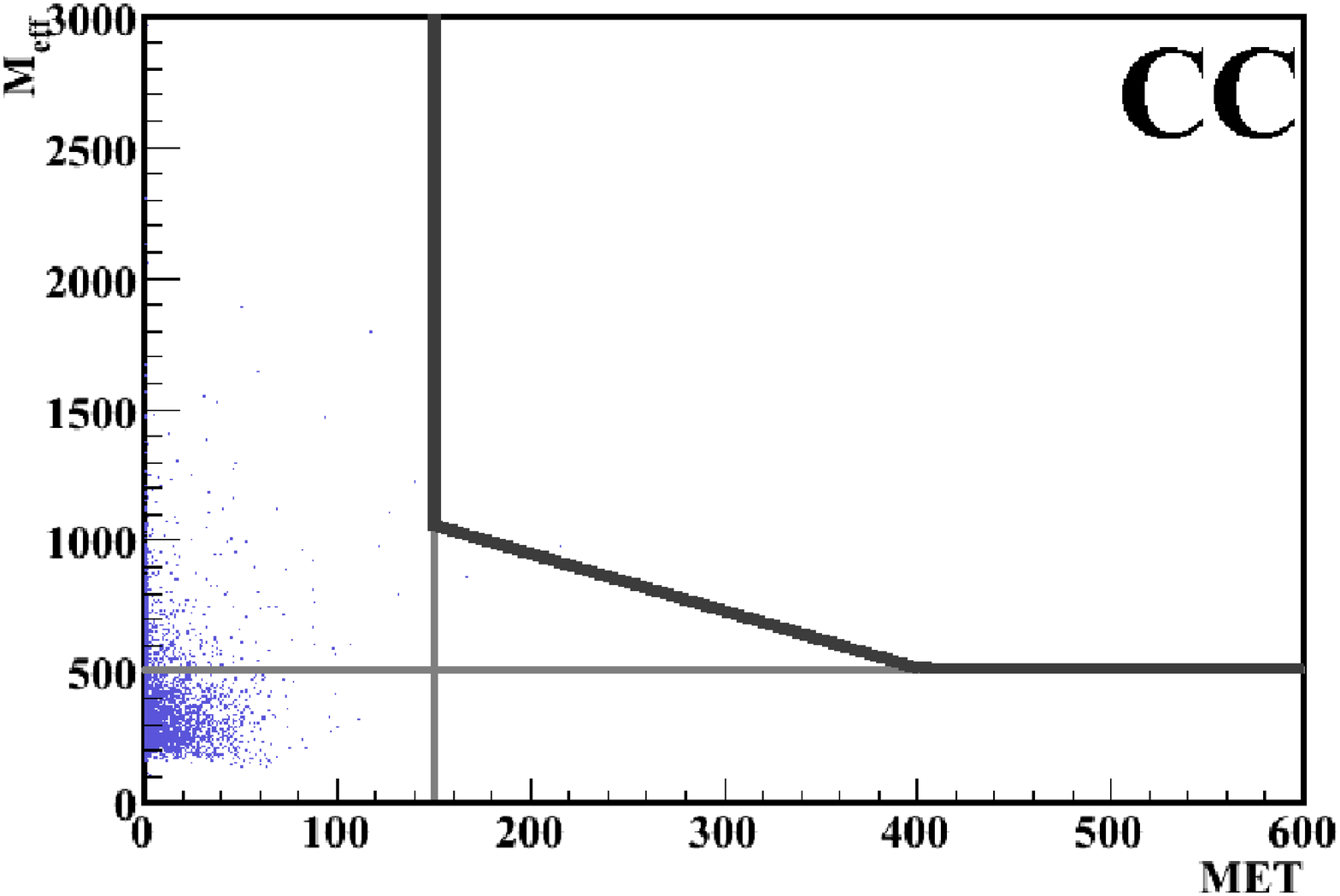}\\
\includegraphics[width=7.9cm]{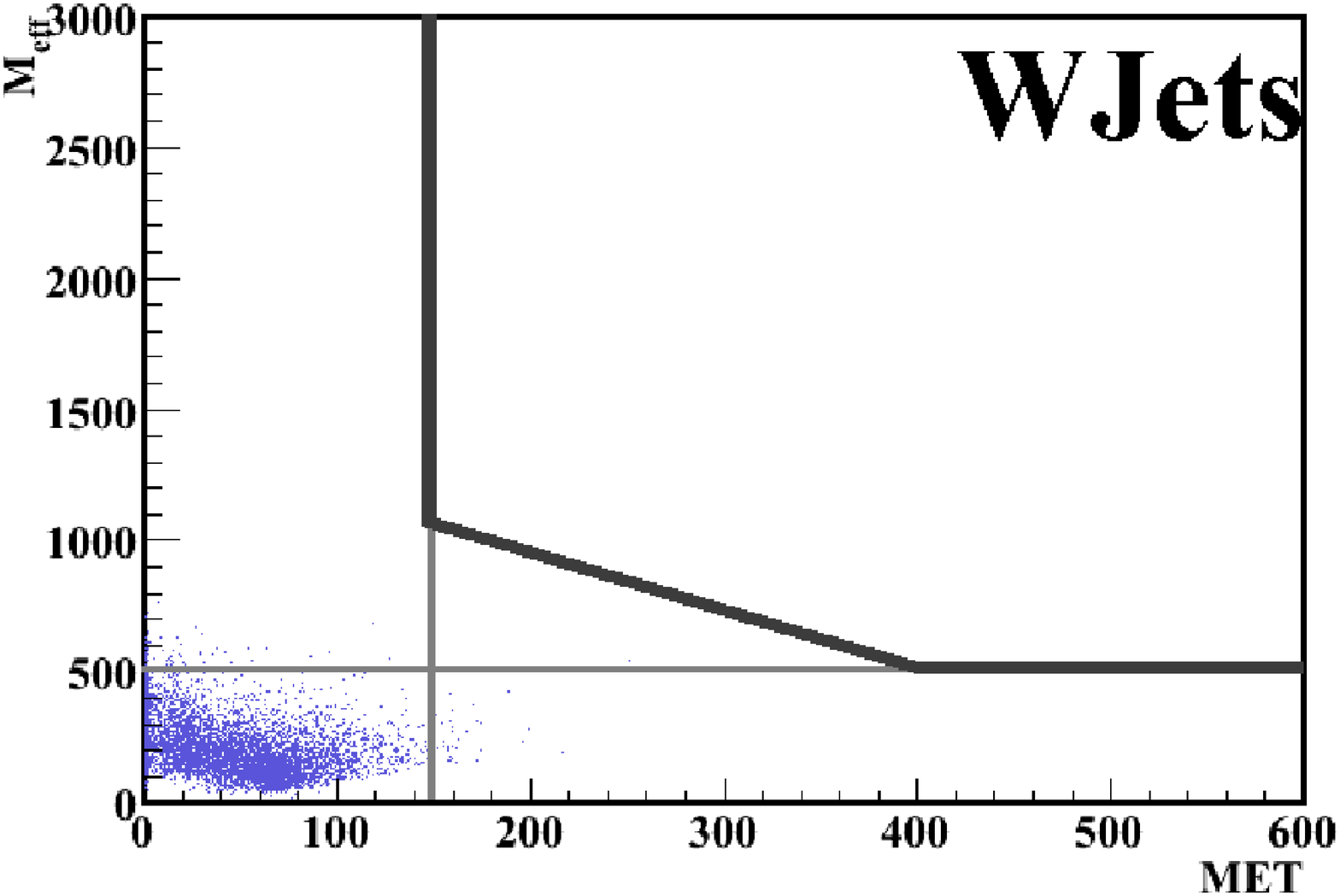}
\includegraphics[width=7.9cm]{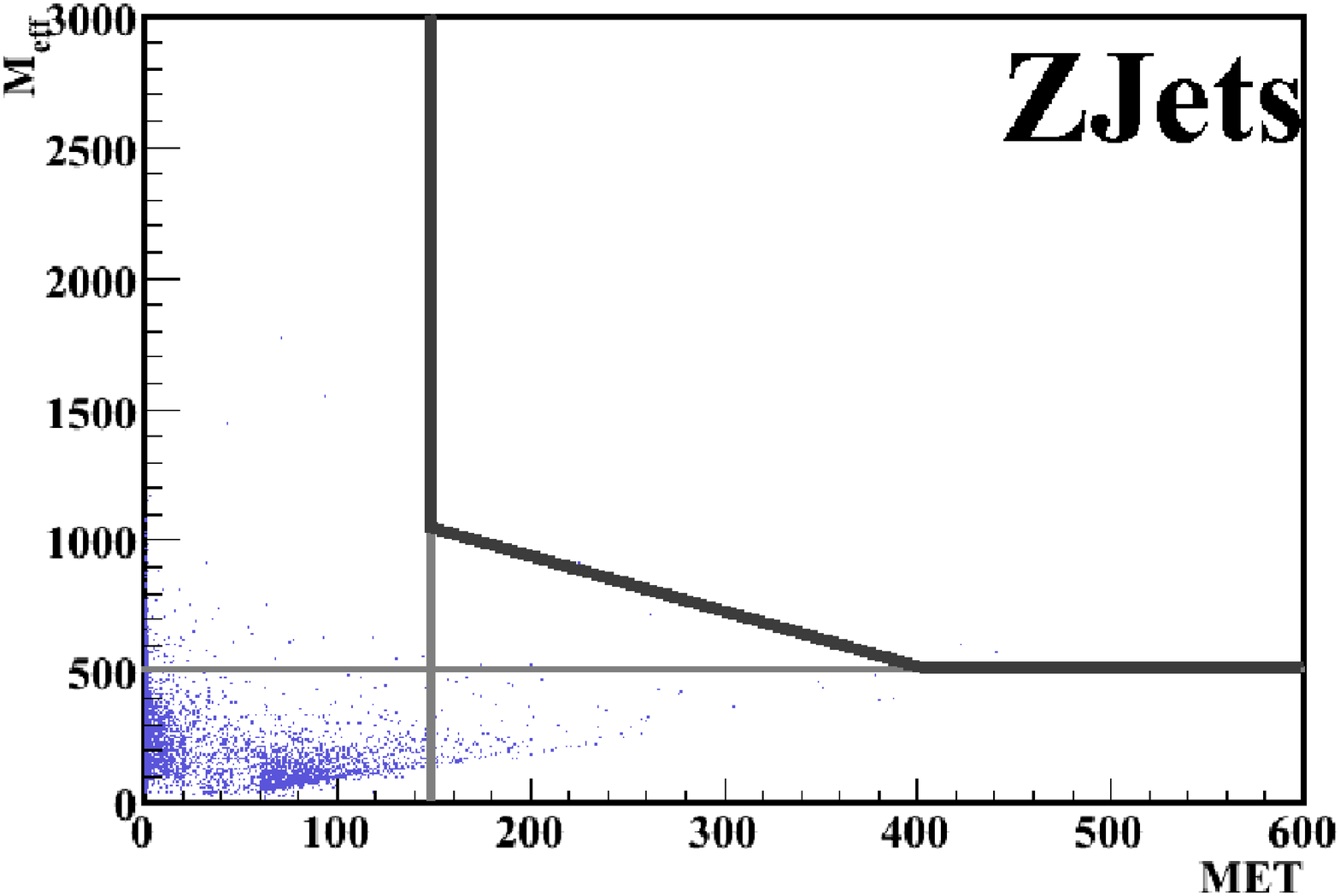}\\
\end{figure}

\begin{figure}[ht]
\caption{Structure of the neural network analysis. The distributions of the backgrounds are not all similar to each other. To avoid a loss of information the analysis has been divided into two parts: The training to different background samples and the training to the outputs of the different neural networks. Twelve variables are given as input for the first step neural networks: $E_T^{\text{miss}}$, $n_{\text{Leptons}}$, $n_{\text{Jets}}$, $n_{\text{B-Jets}}$, $M_{\text{eff}}$,  $cos(\zeta)$, and $p_T$ and $\eta$ of the three leading jets. The final neural network is trained to the outputs of the of the previous three neural networks. As last step a one dimensional cut is done on the final neural network output, which looks well separated after applying the final neural network.\label{FigNNStructure}}
\bigskip
\includegraphics[width=16cm]{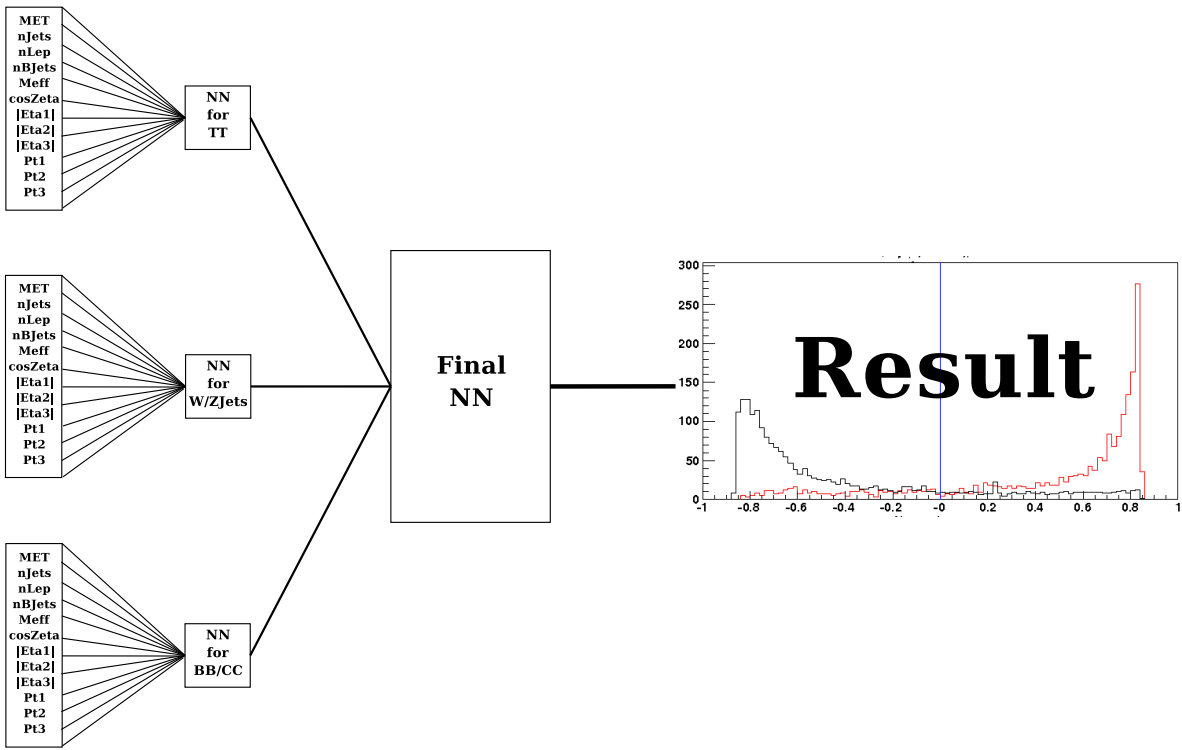}
\end{figure}

\begin{figure}[ht]

\caption{Significance scan in the mSUGRA $m_0$-$m_{1/2}$ plane at $\tan \beta =50$. for an integrated luminosity of 30 fb$^{-1}$. The discovery reach is up to $m_{1/2} = 500$ GeV. The lower right corner is exculded by EWSB and the upper left part is not in accord with WMAP data.\label{FigScan}}
\bigskip
\includegraphics[width=15cm]{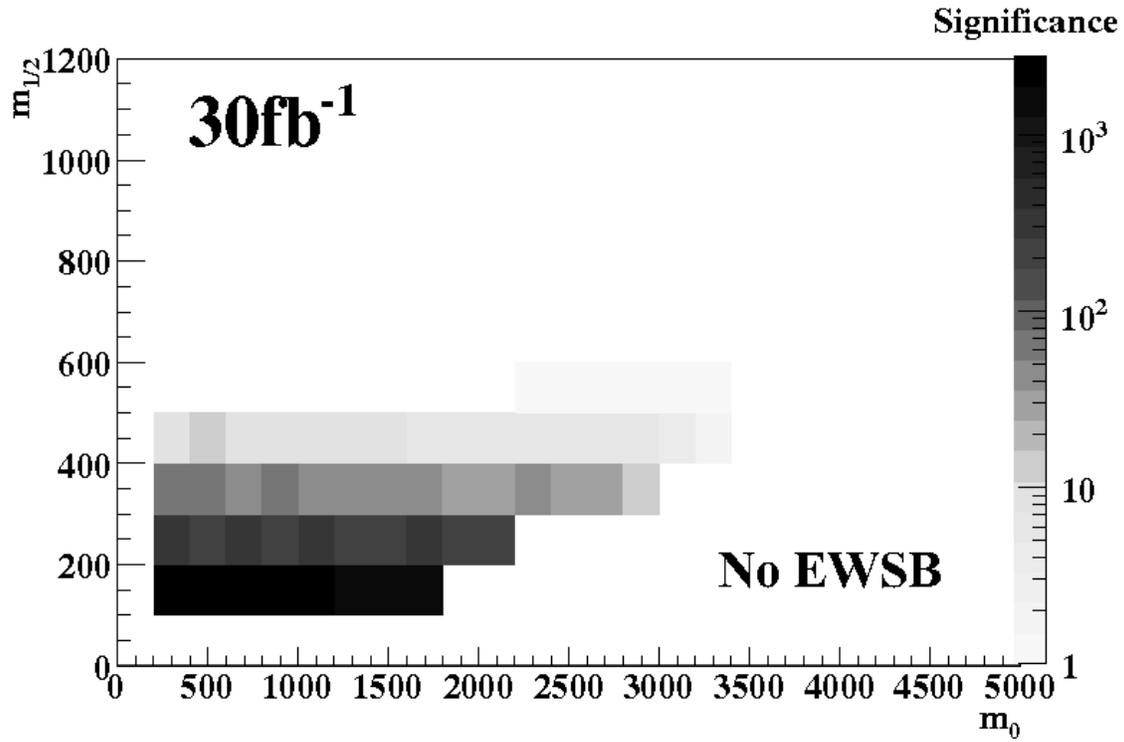}
\end{figure}

\begin{figure}[ht]

\caption{As in Fig. \ref{FigScan}, but for a finer scan in $m_0$ and $m_{1/2}$. Here just for the lower points the significance is high enough to make a discovery. \label{FigScanFPR}}
\bigskip
\includegraphics[width=15cm]{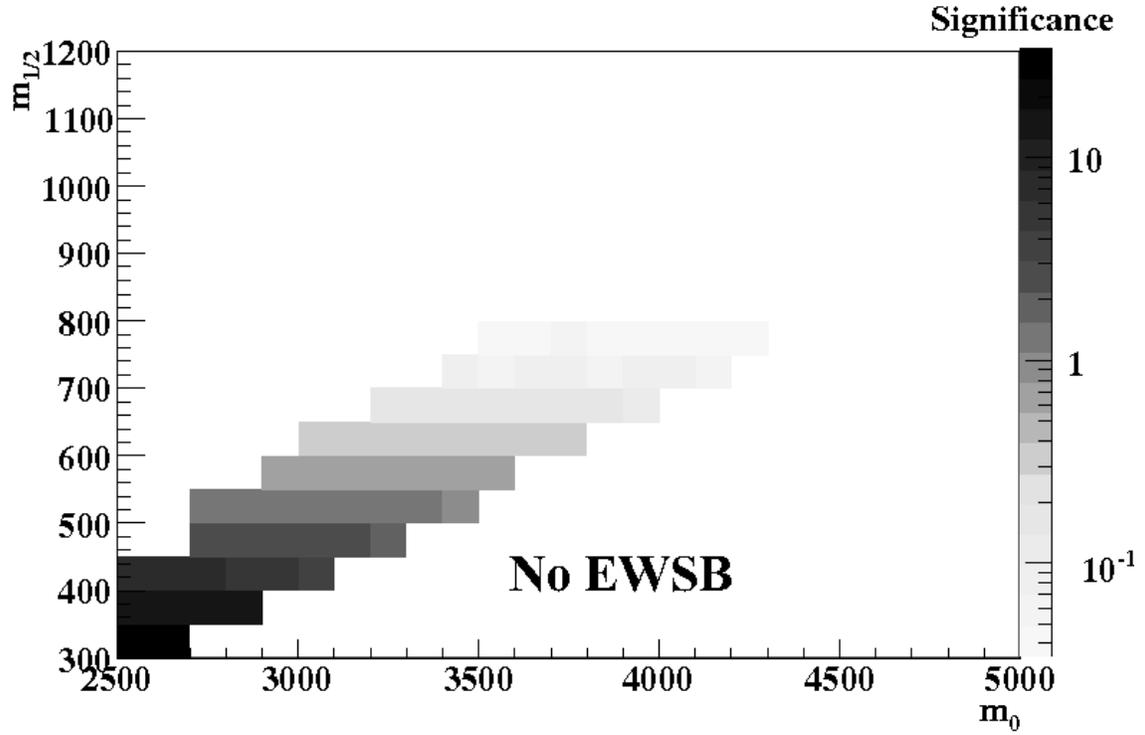}
\end{figure}

The analysis has been done at Monte Carlo level, i.e. excluding detector simulation and reconstruction. The hardware based L1 trigger from CMS has been included. Its trigger table is given in \cite{TDR1} and the resulting trigger efficiency of the signal and background events are all mentioned in table \ref{TabTrig}.\\
The event topology for gluino pair production, which can be assumed from the possible types of cascade decays, is given by:
\begin{equation}
\begin{split}
  	E_T^{\text{miss}} &\text{ high} \\%
	n_{\text{jets}} &\geq 4\\
	M_{\text{eff}} = E_T^{\text{miss}} + \sum_{\text{jets}} p_{T,i} &\text{ high} \\
	n_{\text{leptons}}.&\\ 
\end{split}
\end{equation}
The missing transverse energy ($E_T^{\text{miss}}$) is expected to be high, because the neutralino is expected to escape from the detector. The third condition is caused by the high sfermion masses. In the gluino decay cascades leptons can only be created in decays of neutralinos or charginos. In order to conserve colour charge minimal four jets have to be produced per event. Additionally the effective mass is expected to be high \cite{ATLAS}. That is caused by the fact that SUSY particles are heavy, thus the MET and the transverse momenta of all jets in the events are expected to be high. Cased by the fact, that the gluino decays into electroweak gauginos, also leptons can appear in the event topology. \\
Possible standard model backgrounds for the production of heavy SUSY particles are heavy quark pair production ($t\bar{t}$, $b\bar{b}$ and $c\bar{c}$) and heavy vector boson production (W and Z plus jets). The heavy quark pair production contributes to the missing transverse energy and to the number of jets. Thus the worst background is expected to be $t\bar{t}$, because it is dealing with the heaviest particles, but its cross section is low (less than 1 nb) compared with the other backgrounds. Many orders of magnitude larger cross sections are expected from vector boson production (in the 50 nb range). That is, what makes these backgrounds very dangerous. They contribute to the missing transverse energy, to the number of jets and from the decays of the vector bosons also to the number of leptons.\\
The mSUGRA parameter point our analysis has been tuned to the CMS benchmark point LM9. It is located in the low fermion mass region and its parameter set is given by 
\begin{equation}\label{LM9Param}
 	\lbrace p \rbrace = \lbrace +1,  1450 \text{ GeV}, 175 \text{ GeV}, 0, 50  \rbrace 
\end{equation}
\cite{TDR1}.
The cross section at this parameter point can be calculated using Prospino \cite{Prospino}, and it is 20.8~pb in leading order and 36.7~pb in the next to leading order approximation.\\
The given event topologies for the low fermion mass point LM9 (definition given in (\ref{LM9Param})) are plotted in Fig. \ref{nJetsnLep} and \ref{nJetsMET}. In reality the topologies of the events are distorted by jet and MET misreconstruction. There it can be seen, that the MET is an important variable to distinguish between signal and background events.\\
The separation between signal and background was optimised using neural networks (Neurobayes \cite{Phi-T}). To check the consistency of the neural network the analysis has also been done as cut analysis.\\
In a neural network the variables get different significances. The more the variables contribute to the output, the higher their significances.
Variables, which have been used to train the neural networks and their significances are given in table \ref{TabSigni}. 
The two most significant variables are $M_{\text{eff}}$ and $E_T^{\text{miss}}$. Therefore one can try to separate signal from background in a 2D-scattering plot of $M_{\text{eff}}$ and $E_T^{\text{miss}}$  as shown in Fig. \ref{Fig2DCuts}. The event numbers are shown in the second row of table \ref{TabAllRes}.\\
A neural network is sensitive to statistics, thus it cannot be trained to a sample, where the number of background events is much higher than the one of signal events. Because of the high background cross section precuts have to be applied to the data. The following preselection cuts have been done:
\begin{equation}
\begin{split}
M_{eff} &>500 \text{ GeV} \\
E_T^{miss} &>150 \text{ GeV.}
\end{split} 
\end{equation}
The resulting event numbers are written in the third row of table \ref{TabAllRes}. After applying the preselection the numbers of background events are in the same order of magnitude as the number of signal events, which provides perfect conditions for the training of a neural network.\\
To avoid a loss of information three different neural networks have been trained to different groups of backgrounds. Here backgrounds with high similarities have been taken as one group: WJets and ZJets and $b\bar{b}$ and $c\bar{c}$. $t\bar{t}$ has been chosen as separate group.\\
After the training the problem can be reduced to a problem in a three dimensional variable space of the network outputs. Because the neural networks have been trained separately, the outputs have to be normalised to the different cross sections of the backgrounds.\\ 
To do the final event selection on the strongly correlated outputs of the three neural networks another neural network was trained to the outputs. Thus the problem is in the end reduced to a one-dimensional problem. 
An overview of the neural network analysis is given in Fig. \ref{FigNNStructure}. The obtained one dimensional problem can easily be solved by significance optimisation. The resulting event numbers are listed in table \ref{TabAllRes}.\\
To proof a discovery the significance, which is for a Gaussian distribution defined as 
\begin{equation}
 	\sigma= \frac{S}{\sqrt{\sum_i B_i}},
\end{equation}
where $S$ is the number of signal events and $B_i$ is the number of background events for the background $i$, has to be more than five.\\
With the given results the significance at LM9 can be calculated to be $\sigma \approx 60$. From comparison of the results for the 2D cut method and the neural networks it can be seen that the number of background events is comparable for both cases. What makes the neural network analysis better is the signal efficiency. In the neural network analysis about double the number of events is selected correctly. The fact, that the numbers of background events are in the same order of magnitude after both selections, validates the consistency of the neural network.\\
The analysis has been tuned to one special parameter point. Now it should be checked, how general the analysis is and if it could be applied to other points in mSUGRA parameter space. Thus a scan in the $m_0$-$m_{1/2}$ plane has been done at fixed $\tan \beta$=50. The results are plotted in Fig. \ref{FigScan}.\\
The analysis can be applied for parameter points surrounding LM9. It is sensitive to the gaugino mass parameter $m_{1/2}$ and less sensitive to the scalar mass parameter $m_0$, which is not surprising, because the gluino is a spin 1/2 particle. As shown in Fig. \ref{FigScan} the fermion mass parameters up to 500 GeV and scalar mass parameters up to 3000 GeV are within the reach for an integrated luminosity of 30 fb$^{-1}$, which corresponds to three years LHC running time.\\
In the focus point region (description given in \cite{FPR}) a more detailed parameter scan has been done. The focus point region is interesting for the comparison, due to the fact that in this region the event topology differs from the one at LM9. Here the decay to bottom quarks is not strongly preferred since the scalar mass parameter becomes high enough to give the possibility for a proper coupling to the top quark. The results are shown in Fig. \ref{FigScanFPR}.\\
The analysis works for some lower points in the focus point region, but it does not in the upper area. Here the event topologies get too different from the ones at LM9
\section{Conclusion}
The EGRET experiment motivates the CMS scenario LM9 (\ref{LM9Param}) and the focus point region. 
The Monte Carlo analysis shows that an mSUGRA the significance for a discovery of an LM9 scenario ($\tan\beta=50$) would be about 60 with an integrated luminosity of 1 fb$^{-1}$, which is reached after one year LHC runtime. With an integrated luminosity of 30 fb$^{-1}$ the analysis would cover the low fermion mass region of $(m_0,m_{1/2})=(200-2800,100-400)$.\\
The next steps, are including the detector simulation, which is expected to decrease the significance. Additionally, the event selection should be optimised with real data, as soon as available.
\newpage

\bibliographystyle{apsrev}
\bibliography{MyBib}

@Article{TDR1,
  author =       "CMS~Collaboration",
  title =        "CMS Physics Technical Design Report, Volume II: Physics Performance",
  journal =      "J. Phys. G: Nucl. Part. Phys.",
  volume =       "34",
  pages =        "995-1579",
  year =         "2006",
}

@Article{Unification,
  author =       "W. de Boer and C. Sander",
  title =        "Global electroweak fits and gauge coupling unification",
  journal =      "Physics Letters B",
  volume =       "585",
  pages =        "276-286",
  year =         "2003",
}
@Article{FPR,
  author =       "H. Baer and T. Krupovniackas and S. Profumo and P. Ullino",
  title =        "Model independent approach to focus point supersymmetry: From dark matter to collider searches",
  journal =      "JHEP",
  volume =       "0510",
  pages =        "020",
  year =         "2005",
}

@Article{Prospino,
  author =       "W.~Beenakker and R.~Hopker and M.~Spira and P.~M.~Zerwas",
  title =        "Squark and gluino production at hadron colliders",
  journal =      "Nucl.\ Phys.\ B ",
  volume =       "492",
  pages =        "51",
  year =         "1997",
}

@Article{CERNCour,
  author =       "W.~de~Boer",
  title =        "Do gamma rays reveal our galaxy's dark matter?",
  journal =      "CERN Cour.",
  volume =       "45",
  pages =        "17-19",
  year =         "2005",
}

@Unpublished{TDR2,
  title =        "CMS Physics Technical Design Report, Volume I: Detector Performance and Software",
  author =       "CMS Collaboration",
  year =         "2006",
  note =         "CERN-LHCC-2006-001",
}
@Unpublished{Trilepton,
  title =        "Trilepton final state from neutralino chargino production in mSUGRA",
  author =       "W.~de Boer and I.~Gebauer and M.~Niegel and C.~Sander and M.~Weber and V.~Zhukov and K.~Mazumdar",
  year =         "2006",
  note =         "CERN-CMS-NOTE-2006-113",
}
@Unpublished{Primer,
  title =        "A Supersymmetry Primer",
  author =       "S. P. Martin",
  year =         "2006",
  note =         "arXiv:9709356 [hep-ph]",
}
@Unpublished{Kazakov,
  title =        "Beyond the Standard Model (in Search for Supersymmetry)",
  author =       "D. I. Kazakov",
  year =         "2001",
  note =         "IEKP-KA/2001-1, EKP (Institut f\"ur Experimentelle Kernphysik, Karlsruhe)",
}
@Unpublished{ATLAS,
  title =        "SUSY sensitivity in final states with leptons, jets and missing transverse energy",
  author =       "M. Chiorboli",
  year =         "2007",
  note =         "arXiv:0710.4787 [hep-ph]",
}
@Unpublished{EGRET,
  title =        "Excess of EGRET Galactic Gamma Ray Data interpreted as Dark Matter Annihilation",
  author =       "W. de Boer and M. Herold and C. Sander and V.Zhukov and  A.V. Gladyshev and D.I Kazakov",
  year =         "2007",
  note =         "arXiv:astro-ph/0408272",
}
@Unpublished{PythMan,
  title =        "PYTHIA 6.2 Physics and Manual",
  author =       "T.~Sjostrand and L.~Lonnblad and S.~Mrenna",
  year =         "2001",
  note =         "arXiv:hep-ph/0108264",
}
@Unpublished{Phi-T,
  title =        "A Neural Bayesian Estimator for Conditional Probability Densities",
  author =       "M.~Feindt",
  year =         "2004",
  note =         "arXiv:physics/0402093v1",
}



\begin{thebibliography}{11}
\expandafter\ifx\csname natexlab\endcsname\relax\def\natexlab#1{#1}\fi
\expandafter\ifx\csname bibnamefont\endcsname\relax
  \def\bibnamefont#1{#1}\fi
\expandafter\ifx\csname bibfnamefont\endcsname\relax
  \def\bibfnamefont#1{#1}\fi
\expandafter\ifx\csname citenamefont\endcsname\relax
  \def\citenamefont#1{#1}\fi
\expandafter\ifx\csname url\endcsname\relax
  \def\url#1{\texttt{#1}}\fi
\expandafter\ifx\csname urlprefix\endcsname\relax\def\urlprefix{URL }\fi
\providecommand{\bibinfo}[2]{#2}
\providecommand{\eprint}[2][]{\url{#2}}

\bibitem[{\citenamefont{de~Boer and Sander}(2003)}]{Unification}
\bibinfo{author}{\bibfnamefont{W.}~\bibnamefont{de~Boer}} \bibnamefont{and}
  \bibinfo{author}{\bibfnamefont{C.}~\bibnamefont{Sander}},
  \bibinfo{journal}{Physics Letters B} \textbf{\bibinfo{volume}{585}},
  \bibinfo{pages}{276} (\bibinfo{year}{2003}).

\bibitem[{\citenamefont{de~Boer et~al.}(2007)\citenamefont{de~Boer, Herold,
  Sander, V.Zhukov, Gladyshev, and Kazakov}}]{EGRET}
\bibinfo{author}{\bibfnamefont{W.}~\bibnamefont{de~Boer}},
  \bibinfo{author}{\bibfnamefont{M.}~\bibnamefont{Herold}},
  \bibinfo{author}{\bibfnamefont{C.}~\bibnamefont{Sander}},
  \bibinfo{author}{\bibnamefont{V.Zhukov}},
  \bibinfo{author}{\bibfnamefont{A.}~\bibnamefont{Gladyshev}},
  \bibnamefont{and} \bibinfo{author}{\bibfnamefont{D.}~\bibnamefont{Kazakov}}
  (\bibinfo{year}{2007}), \bibinfo{note}{arXiv:astro-ph/0408272}.

\bibitem[{\citenamefont{Martin}(2006)}]{Primer}
\bibinfo{author}{\bibfnamefont{S.~P.} \bibnamefont{Martin}}
  (\bibinfo{year}{2006}), \bibinfo{note}{arXiv:9709356 [hep-ph]}.

\bibitem[{\citenamefont{de~Boer}(2005)}]{CERNCour}
\bibinfo{author}{\bibfnamefont{W.}~\bibnamefont{de~Boer}},
  \bibinfo{journal}{CERN Cour.} \textbf{\bibinfo{volume}{45}},
  \bibinfo{pages}{17} (\bibinfo{year}{2005}).

\bibitem[{\citenamefont{Baer et~al.}(2005)\citenamefont{Baer, Krupovniackas,
  Profumo, and Ullino}}]{FPR}
\bibinfo{author}{\bibfnamefont{H.}~\bibnamefont{Baer}},
  \bibinfo{author}{\bibfnamefont{T.}~\bibnamefont{Krupovniackas}},
  \bibinfo{author}{\bibfnamefont{S.}~\bibnamefont{Profumo}}, \bibnamefont{and}
  \bibinfo{author}{\bibfnamefont{P.}~\bibnamefont{Ullino}},
  \bibinfo{journal}{JHEP} \textbf{\bibinfo{volume}{0510}}, \bibinfo{pages}{020}
  (\bibinfo{year}{2005}).

\bibitem[{\citenamefont{de~Boer et~al.}(2006)\citenamefont{de~Boer, Gebauer,
  Niegel, Sander, Weber, Zhukov, and Mazumdar}}]{Trilepton}
\bibinfo{author}{\bibfnamefont{W.}~\bibnamefont{de~Boer}},
  \bibinfo{author}{\bibfnamefont{I.}~\bibnamefont{Gebauer}},
  \bibinfo{author}{\bibfnamefont{M.}~\bibnamefont{Niegel}},
  \bibinfo{author}{\bibfnamefont{C.}~\bibnamefont{Sander}},
  \bibinfo{author}{\bibfnamefont{M.}~\bibnamefont{Weber}},
  \bibinfo{author}{\bibfnamefont{V.}~\bibnamefont{Zhukov}}, \bibnamefont{and}
  \bibinfo{author}{\bibfnamefont{K.}~\bibnamefont{Mazumdar}}
  (\bibinfo{year}{2006}), \bibinfo{note}{cERN-CMS-NOTE-2006-113}.

\bibitem[{\citenamefont{Collaboration}(2006)}]{TDR1}
\bibinfo{author}{\bibfnamefont{C.}~\bibnamefont{Collaboration}},
  \bibinfo{journal}{J. Phys. G: Nucl. Part. Phys.}
  \textbf{\bibinfo{volume}{34}}, \bibinfo{pages}{995} (\bibinfo{year}{2006}).

\bibitem[{\citenamefont{Chiorboli}(2007)}]{ATLAS}
\bibinfo{author}{\bibfnamefont{M.}~\bibnamefont{Chiorboli}}
  (\bibinfo{year}{2007}), \bibinfo{note}{arXiv:0710.4787 [hep-ph]}.

\bibitem[{\citenamefont{Beenakker et~al.}(1997)\citenamefont{Beenakker, Hopker,
  Spira, and Zerwas}}]{Prospino}
\bibinfo{author}{\bibfnamefont{W.}~\bibnamefont{Beenakker}},
  \bibinfo{author}{\bibfnamefont{R.}~\bibnamefont{Hopker}},
  \bibinfo{author}{\bibfnamefont{M.}~\bibnamefont{Spira}}, \bibnamefont{and}
  \bibinfo{author}{\bibfnamefont{P.~M.} \bibnamefont{Zerwas}},
  \bibinfo{journal}{Nucl.\ Phys.\ B} \textbf{\bibinfo{volume}{492}},
  \bibinfo{pages}{51} (\bibinfo{year}{1997}).

\bibitem[{\citenamefont{Feindt}(2004)}]{Phi-T}
\bibinfo{author}{\bibfnamefont{M.}~\bibnamefont{Feindt}}
  (\bibinfo{year}{2004}), \bibinfo{note}{arXiv:physics/0402093v1}.

\bibitem[{\citenamefont{Sjostrand et~al.}(2001)\citenamefont{Sjostrand,
  Lonnblad, and Mrenna}}]{PythMan}
\bibinfo{author}{\bibfnamefont{T.}~\bibnamefont{Sjostrand}},
  \bibinfo{author}{\bibfnamefont{L.}~\bibnamefont{Lonnblad}}, \bibnamefont{and}
  \bibinfo{author}{\bibfnamefont{S.}~\bibnamefont{Mrenna}}
  (\bibinfo{year}{2001}), \bibinfo{note}{arXiv:hep-ph/0108264}.

\end{thebibliography}


@ARTICLE{feyn54,
   author = "R. P. Feynman",
   year = "1954",
   journal = "Phys.\ Rev.",
   volume = "94",
   pages = "262"
}

@ARTICLE{epr,
   author = "A. Einstein and B. Podolsky and N. Rosen",
   year = "1935",
   journal = "Phys.\ Rev.",
   volume = "47",
   pages = "777"
}

@MISC{witten2001,
   author = "Edward Witten",
   eprint = "hep-th/0106109"
}

\newpage

\newpage

\newpage

\newpage
\newpage
\newpage

\newpage
%
\newpage
%
\newpage
\newpage
\newpage
\newpage

\newpage
\begin{table}[ht]
\caption{Evolution of event numbers during the analysis\label{TabTrig} \label{TabPreRes} for 1 fb$^{-1}$. First the event numbers without any analysis tool are plotted. The $\hat{p}_T$ (definition given in \cite{PythMan}) cuts chosen for the generation are the following: 50 GeV for ZJets, 85 GeV for WJets, and 180 GeV for $b\bar{b}$ and $c\bar{c}$. In the first step the CMS L1 trigger is included, where WJets and ZJets already get reduced very well, while 87\% of the signal events pass the trigger. Backgrounds that also pass the trigger nearly untouched are $t\bar{t}$ and $b\bar{b}$. After applying the preselection cuts the number of background events is in the same order of magnitude like the number of signal events. Also the number of $t\bar{t}$ is just about four times bigger than the number of signal events. After applying the final selection the final significance can be calculated to be $\sigma=54$.}
\bigskip
\begin{tabular}{|c|c|c|c|c|}
 \hline
& Before Selections & After L1 Trigger& After Preselection & After Final Selection\\
\hline
\hline
Gluinos Pairs at LM9 &1.5 $\cdot$ 10$^4$ &1.2 $\cdot$ 10$^4$& 5.0$\cdot$ 10$^3$& 3.7$\cdot$ 10$^3$\\
WJets & 1.3 $\cdot$ 10$^7$&1.58 $\cdot$ 10$^6$&1.3 $\cdot$ 10$^3$& 1.0$\cdot$ 10$^1$\\
ZJets &  1.15 $\cdot$ 10$^7$&8.05 $\cdot$ 10$^5$& 4.7 $\cdot$ 10$^3$& 5.0$\cdot$ 10$^2$\\
$t \bar{t}$ & 6.2 $\cdot$ 10$^5$ &5.75 $\cdot$ 10$^5$& 2.0 $\cdot$ 10$^4$& 2.9$\cdot$ 10$^3$ \\
$b\bar{b}$ &8.00 $\cdot$ 10$^6$ &5.61 $\cdot$ 10$^6$& 4.8 $\cdot$ 10$^3$& 1.2$\cdot$ 10$^3$\\
$c\bar{c}$ & 8.05 $\cdot$ 10$^6$&5.48 $\cdot$ 10$^6$& 1.3 $\cdot$ 10$^3$& 3.0$\cdot$ 10$^2$\\
\hline
\end{tabular}
\end{table}
\newpage
\begin{table}[ht]
\caption{Significances of the variables used for the neural network training.\label{TabSigni} The three jets chosen are the three hardest jets in the event and $\zeta$ is the angle between the missing transverse momentum and the transverse momentum of the hardest jet.}
\bigskip
\begin{tabular}{|c|c|c|c|}
 \hline
&  TT-Network &WZ-Network& BB/CC-Network\\
\hline
\hline
$M_{\text{eff}}$ & 60.9\%&73.0\%&28.8\%\\
$M_{T}^{miss}$ &27.2\%&16.3\%&24.3\%\\
$\cos(\zeta)$ &31.0\%&46.5\%&52.3\%\\
$P_T$ of highest $p_T$ jet & 30.0\% &59.0\%&50.8\%\\
$P_T$ of second jet &50.0\%&68.7\%&47.1\%\\
$P_T$ of third jet &56.1\%&77.9\%&27.0\%\\
$\eta$ of highest $p_T$ jet& 15.8\%&20.2\%&8.8\%\\
$\eta$ of second jet& 14.8\%&33.0\%&10.1\%\\
$\eta$ of third jet&15.7\%&29.7\%&15.0\%\\
Number of jets&36.1\%&64.3\%&62.0\%\\
Number of B-jets& 1.6\%&41.8\%&7.3\%\\
Number of leptons&16.9\%&65.2\%&27.3\%\\
\hline
\end{tabular}
\end{table}
\newpage
\begin{table}[ht]
\caption{Results of the two dimensional cuts and the neural network.\label{TabAllRes} The outputs of the neural network analysis and the 2D analysis are in the same order of magnitude, which validates the neural network analysis. For the neural network analysis the signal efficiency is higher and the $t\bar{t}$ and the $b\bar{b}$ as well as the ZJets selection is better, while the $c\bar{c}$ and WJets selections are a bit worse in the neural network analysis. That is caused by the fact, that these backgrounds are trained together with other backgrounds. So if one background separation gets better the other will get a bit worse. But in case of our analysis the avail is larger than the loss.}
\bigskip
\begin{tabular}{|c|c|c|}
 \hline
&  2D Cuts & Neural Network\\
\hline
\hline
Gluinos Pairs at LM9 & 3.2 $\cdot$ 10$^3$ & 3.7 $\cdot$ 10$^3$\\
WJets &7.0 $\cdot$ 10$^1$ & 1.0 $\cdot$ 10$^1$\\
ZJets & 9.0 $\cdot$ 10$^2$ & 5.0 $\cdot$ 10$^2$\\
$t \bar{t}$ & 3.1 $\cdot$ 10$^3$& 2.9 $\cdot$ 10$^3$\\
$b\bar{b}$ & 2.9 $\cdot$ 10$^3$& 1.2 $\cdot$ 10$^3$ \\
$c\bar{c}$ &9.0 $\cdot$ 10$^2$ & 3.0 $\cdot$ 10$^2$\\
\hline
\end{tabular}
\end{table}
\newpage
\textbf{Eva Ziebarth}\\
\\
\textbf{Address:}\\
Institut fuer Experimentelle Kernphysik\\
Universitaet Karlsruhe (TH)\\
Wolfgang-Gaede-Str. 1\\
Gebaeude 30.23\\
D-76131 Karlsruhe\\
Germany\\
\\
\textbf{Phone:} \\
+49-721-6087577\\
\\
\textbf{E-Mail:}\\
ziebarth@cern.ch

\end{document}